\documentclass[%
 letter
 amsmath,amssymb,
 aps,
 prd
]{revtex4-2}

\usepackage{graphicx}
\usepackage{graphicx}
\usepackage{dcolumn}
\usepackage{bm}

\usepackage{url}
\usepackage{parskip}
\usepackage[ruled,lined]{algorithm2e}
\usepackage{algpseudocode} 
\usepackage{subcaption}
\usepackage{amsmath}
\usepackage{booktabs}
\usepackage{multirow}
\usepackage{hyperref}      

\newcommand{\fastcal}{Fast Calorimeter Simulation Challenge 2022}
\begin{document}
\preprint{APS/123-QED}
\title{CaloGraph: Graph-based diffusion model for fast shower generation in calorimeters with irregular geometry}

\author{Dmitrii Kobylianskii}
\email{dmitry.kobylyansky@weizmann.ac.il}
\author{Nathalie Soybelman}%
\author{Etienne Dreyer}%
\author{Eilam Gross}%
\affiliation{%
 Weizmann Institute of Science, Israel
}%
\date{\today}

\begin{abstract}
Denoising diffusion models have gained prominence in various generative tasks, prompting their exploration for the generation of calorimeter responses. Given the computational challenges posed by detector simulations in high-energy physics experiments, the necessity to explore new machine-learning-based approaches is evident. This study introduces a novel graph-based diffusion model designed specifically for rapid calorimeter simulations. The methodology is particularly well-suited for low-granularity detectors featuring irregular geometries. We apply this model to the ATLAS dataset published in the context of the \fastcal, marking the first application of a graph diffusion model in the field of particle physics.
\end{abstract}

\maketitle
\section{Introduction}
Simulation plays an essential role in interpreting collision data from the Large Hadron Collider (LHC) experiments and testing alignment with theoretical predictions. The unique set of challenges entailed in simulating collision data, including high-dimensional feature space and lack of tractable likelihood models, have inspired a range of deep learning solutions \cite{mlandlhcgen,deepgendet}. In particular, for simulating particle interactions in the detector, the core challenge is limited computational resources, dominated by the extreme detail needed to model particle showers in the calorimeter. Here, the traditional approach of Monte Carlo simulation based on \textsc{Geant4}~\cite{geant} is robust but highly resource intensive -- occupying the largest fraction of time in the ATLAS simulation chain \cite{SOFT-2010-01}. In future high-luminosity LHC runs, calorimeter simulations will need to cope with an order of magnitude higher data rate, potentially becoming the limiting factor for physics analysis without significant progress in the field \cite{hl-lhc}.

Many efforts have been employed in order to speed up calorimeter simulations significantly. While fast parameterized shower models have been successfully deployed at LHC experiments~\cite{atlasfastcalosim,cmsfastcalosim}, they are limited in accuracy. More recently, the emergence of deep generative models has led to their great popularity and potential in tackling this task. The first generative model applied to calorimeter simulations was \textsc{CaloGAN}~\cite{calogan}. It represented the calorimeter as a 3D image and used a Generative Adversarial Network (GAN) to generate them. Building on the success of this work, GANs were already implemented in the ATLAS fast simulation \textsc{AtlFast3}~\cite{ATLAS:2021pzo}. 

New developments in the field of generative models bringing new models to the market triggered a multitude of new developments for calorimeter simulations~\cite{caloclouds,caloclouds2,calodiffusion,caloflow_ds1,caloinn,caloscore,caloscorev2,caloshowergan,supercalo,inductive_caloflow}. The \textit{\fastcal}~\cite{ds1,ds2,ds3} was developed to provide a framework for a comparison of all ongoing efforts. Normalizing flows (NFs) and invertible neural networks (INNs), as well as various diffusion-based models, showed promising results, while often exhibiting a compromise between accuracy and speed. Beyond the choice of network architecture, the data representation poses an important aspect as well. While, for example, \textsc{CaloScore}~\cite{caloscore,caloscorev2} and \textsc{CaloDiffusion}~\cite{calodiffusion} rely on image representation, \textsc{CaloClouds}~\cite{caloclouds,caloclouds2} and \textsc{CaloFlow}~\cite{inductive_caloflow} utilize point clouds. Point clouds offer a natural representation of calorimeters and particle physics data in general. Generative models of point clouds proved their effectiveness in the task of jet simulations~\cite{jetben,jetepic,epicGAN,pcdroid,pointcloudtrfo}, where each point represents a jet constituent. For calorimeter simulations, points represent energy depositions. Such a representation is independent of the underlying geometry and, therefore, computationally more efficient than a sparse 3d image. However, voxelization needs to be applied to convert the point cloud into detector cells, which can introduce a bias and affect the performance. This method is, therefore, mostly suitable for high-granularity calorimeters, for example, at the proposed CLIC detector~\cite{clic}.
\begin{figure}[t]
    \centering
    \includegraphics[trim={0 15cm 0 10cm},width=0.7\linewidth]{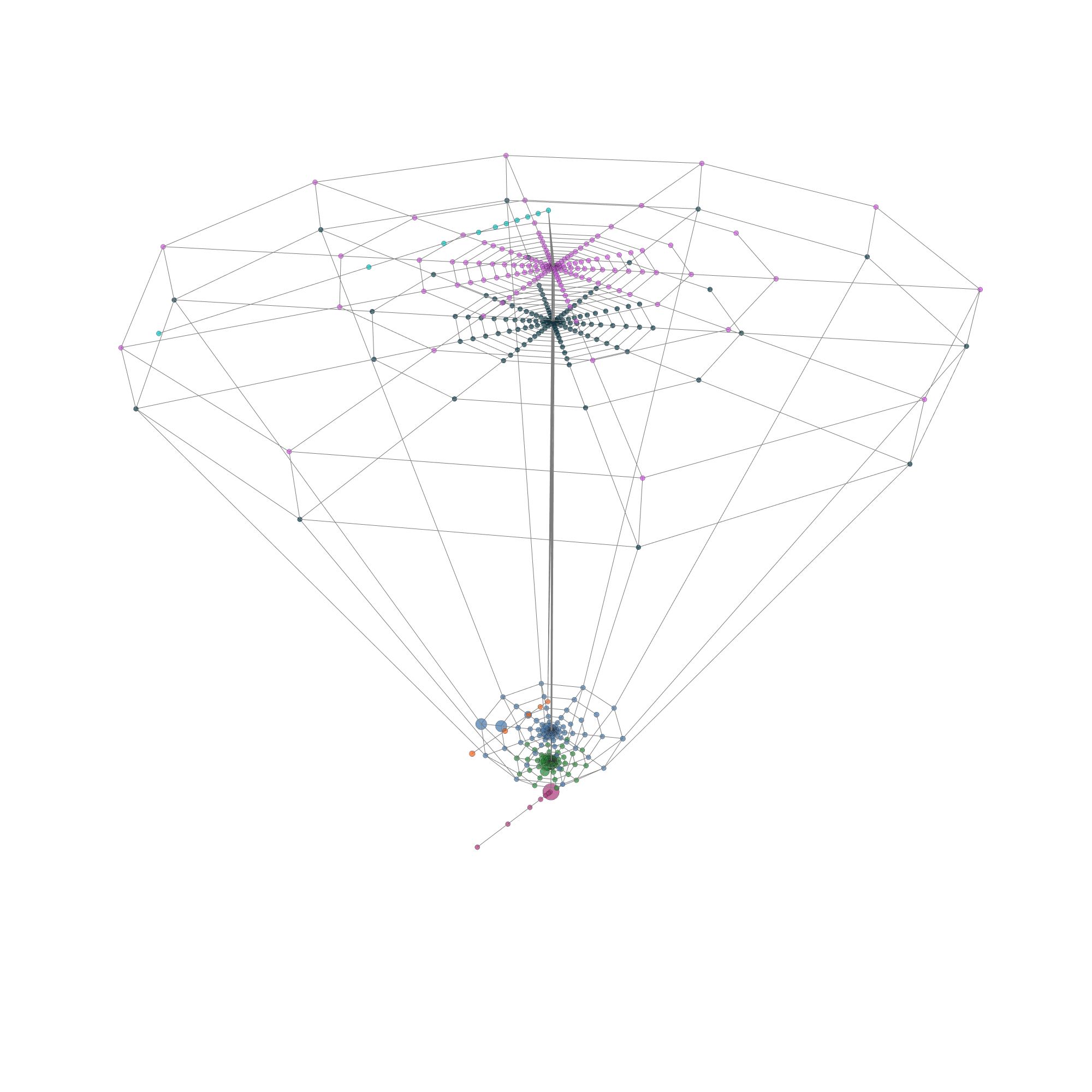}
    \caption{Event display from the pion dataset in graph form. The nodes represent voxels, and their size relates monotonically to the deposited energy. Each colour represents a different calorimeter layer.}
    \label{fig:pion-graph}
\end{figure}

In this work, we propose \textsc{CaloGraph}: a denoising diffusion model that represents calorimeter data as a graph, as shown in Fig.~\ref{fig:pion-graph}. Graph neural networks (GNNs) showed wide success in HEP applications~\cite{Shlomi_2020}, including clustering tracking detector hits into tracks~\cite{graphtrack,graphtrack2,graphtrack3} or classifying jets based on constituents~\cite{jettag,jettag2,GN1}. Particularly, graph representations of calorimeters are used for the reconstruction of particle flow candidates~\cite{CMSpflow,MLPF,HGpflow}. Until now, their use for generative tasks remained unexplored.
The motivation for a graph representation of the calorimeter is similar to the one for point clouds, with the additional advantage of edges incorporating relations between neighbouring objects and allowing information transfer. Furthermore, unlike the point cloud, the graph structure is fixed, and nodes directly represent detector cells, avoiding the need for voxelization and associated performance losses.
However, a large number of edges will lead to a significant memory need, posing a limitation to this method for high-granularity detectors. 
In most inference tasks with GNNs, the number of nodes is predefined by the input graph, whereas in our case, nodes with nonzero energy are predicted by the generation task and thus not known in advance.
It is therefore necessary to define a fixed initial graph that sets an upper limit on the number of nodes with nonzero energy. We therefore present results for the ATLAS-like dataset 1 of the calorimeter challenge~\cite{ds1}.

We note that applications of graph-based diffusion have thus far been restricted to generation of molecules, proteins, and materials~\cite{graphdiffreview}. Typically in these cases, the network is used to predict the graph adjacency matrix and also node and edge features in some cases \cite{molecules}. In our application, we take the graph structure as given and only predict node features with the diffusion model. 

\section{Dataset}

\textsc{CaloGraph} was trained and evaluated using Dataset 1 from the \fastcal~\cite{ds1}. This dataset comprises \textsc{Geant4}-simulated particle showers in a model of the ATLAS calorimeter and was used to train \textsc{AtlFast3}~\cite{ATLAS:2021pzo}. The dataset comprises two sets of calorimeter showers—one for photons and one for pions. 

The calorimeter consists of 533 voxels and 7 layers in the case of the pion dataset and 368 voxels and 5 layers for photons. The voxelization leads to an irregular distribution of radial and angular bins $N_r \times N_\alpha$ among the layers given as follows:
\begin{align}
    \text{pions} & \qquad 8\times1,\;10\times 10,\;10\times 10,\; 5\times 1,\; 5\times 1,\; 15\times 10,\;16\times 10,\; 10\times 1\nonumber\\
    \text{photons} & \qquad 8\times1,\;16\times 10,\;19\times 10,\; 5\times 1,\; 5\times 1 
\end{align}
Photon and pion energies range from 256 MeV to 4.2 TeV, increasing in powers of two. For each energy level, 10,000 showers are provided, with a reduced amount for very high energies. In total 242,000 (241,600) showers are provided for the photon (pion) dataset.
\subsection{Graph creation}
Each calorimeter shower is represented by a graph in the following way. An example from the pion sample is shown in Fig.~\ref{fig:pion-graph}. The graph nodes depict the calorimeter voxels with the fixed coordinates $\{\eta_i, \phi_i, \text{layer}_i\}$ as input features and the deposited energy $E_i$ as ground truth.
As shown, the voxels comprise rings of equal $R_i = \sqrt{\phi_i^2 + \eta_i^2}$ which are used to determine the graph edges. Each voxel is connected to its two neighbouring voxels on the same ring and up to two nearest neighbors on concentric rings.
Connections between layers occur for voxels located in the outermost and innermost rings in terms of $R$. 
In layers without angular binning, the rings each correspond to a single voxel and are connected concentrically, as depicted by straight lines in Fig.~\ref{fig:pion-graph}. The innermost voxels in these layers are connected to the innermost ring of voxels in the layers below and above, where available. In total, we have 1478 (2210) edges for the photon (pion) graphs.

\subsection{Data preprocessing}
Before the network operates on the graph, a normalization process is applied to the node features. The $\eta$ and $\phi$ coordinates are normalized to achieve a zero mean and a standard deviation of 1. The energy normalization process is illustrated in the pseudo-code of Algorithm~\ref{alg:normalize}. The deposited energies in the voxels $E_{i}$ are normalized relative to the incoming energy $E_{inc}$ with an additional factor $f = 3.1$ for photons and $f = 6.4$ for pions. This factor is essential to account for scenarios where $E_{inc}$ is less than $E_{i}$. Following this, a logit transform is executed with $\alpha=10^{-6}$. Finally, the resulting values undergo further normalization to obtain a zero mean and a standard deviation of 1.  The incoming energy $E_{inc}$ undergoes min-max normalization to be in the range $[0, 1]$.

\begin{algorithm}[b]
\caption{Deposited energy normalization procedure}\label{alg:normalize}
\begin{algorithmic}[1]
\Procedure{Normalize}{$E_{i}, E_{inc}$}
\State $E_{i}' \gets \frac{E_{i}}{f \cdot E_{inc}}$
\State $E_{i}' \gets \alpha + (1 - 2\alpha) E_{i}'$
\State $E_{i}' \gets \ln{\frac{E_{i}'}{1 - E_{i}'}}$
\State $E_{i}' \gets \frac{E_{i}' - \mu}{\sigma}$
\State \textbf{return} $E_{i}'$ \Comment{Return normalized energy}
\EndProcedure
\end{algorithmic}
\end{algorithm}
\section{Model description}

\textsc{CaloGraph} is a score-based graph diffusion model, where a diffusion process is designed to perturb the original node features $x$ slowly with Gaussian noise addition. The neural network learns to undo the noising process by 
estimating the amount of noise added $\epsilon_t$ and solving the corresponding differential equation.
In this way, we can start from pure noise $x_T \sim \mathcal{N}(0, \mathbb{I})$ and sequentially denoise it to sample from the original data distribution $x_0 \sim p_{data}$. 

The diffusion formulation of \textsc{CaloGraph} follows the DDIM approach~\cite{ddim} closely. We take as input a noised graph with disturbed energy $E_{t} = \sqrt{\bar{\alpha}_t}E_{0} + \sqrt{1 - \bar{\alpha}_t}\epsilon$, where $t$ is the diffusion time step, $\epsilon \sim \mathcal{N}(0, \mathbb{I})$ and $\bar{\alpha}_t$ are given by a cosine noise schedule adopted from~\cite{improved_diffu}. During training, the GNN learns to predict the added noise~$\epsilon$ conditioned on the voxel coordinates $C$ and the incoming particle energy $E_{inc}$. The loss is defined as:
\begin{equation}
    L = \mathbb{E}_{t, E_0, \epsilon} \big[\left\|\epsilon_\theta(E_t; t, C, E_{inc}) - \epsilon \right\|^2\big]
\end{equation}
During the inference process, we start with the graph, where the deposited energies are given by pure noise $E_T \sim \mathcal{N}(0, \mathbb{I})$. Then, we use our network and the PNDM sampler from~\cite{liu_pseudo_2022} to estimate the noise needed to be removed to iteratively denoise the deposited energy.
\begin{figure}[t]
    \centering
    \includegraphics[width=\textwidth]{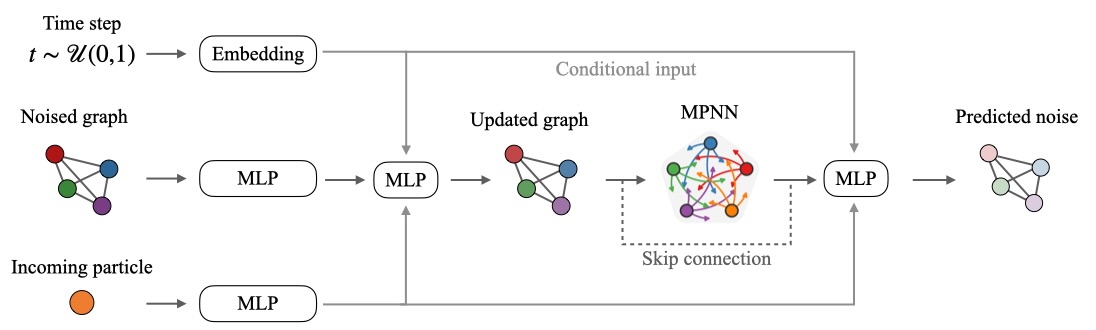}
    \caption{Architecture of \textsc{CaloGraph}. The input noised graph is updated with the embedded conditional input comprised of the sampled time step and the incoming particle energy. Message-passing is applied on the updated graph, and the output is combined with the conditional input and the pre-message-passing node features to predict the noise.}
    \label{fig:calograph}
\end{figure}
The \textsc{CaloGraph} architecture is depicted in Fig.~\ref{fig:calograph}. The network takes as input the noised target graph with the node features, including the cell positions ($\eta_i, \phi_i$, layer$_i$) and the noised cell energy $E_t$ as described above. Initially, the cell layer undergoes an update through a learnable embedding. Following this, both position and noised energy are processed through a multi-layer perceptron (MLP). The updated node features are then combined with the conditional input, comprising the embedded, uniformly sampled time step and the incoming particle energy updated via another MLP. The resulting combined output is passed through another MLP, giving the updated graph.
Subsequently, four rounds of message-passing neural networks (MPNN)~\cite{mpnn} are applied. The output of the message passing is combined through a skip connection with the input to the MPNN and the conditional input. The resulting node representation vectors are passed through the final MLP to predict the noise.
We trained separate models for the photon and pion samples using the same architecture and hyperparameters shown in Tab.~\ref{tab:hyperparams}, without dedicated optimization. The models were implemented using \textsc{PyTorch} and DGL (Deep Graph Library)~\cite{dgl}.
\begin{table}[b]
    \centering
    \begin{tabular*}{0.6\linewidth}{@{\extracolsep{\fill}}lr}
    \toprule
    \multicolumn{2}{@{}l}{\textbf{Hyperparameters}}\\\hline
    batch size & 200\\
    optimizer & AdamW\\
    lr & $10^{-3}$\\
    \# of epochs & 700\\
    \# of time steps & 50\\
    \toprule
    \textbf{Network sub-parts} &  \textbf{Parameters}\\
    Embedding part & 155 710\\
    MPNN (4 rounds) & 519 750\\
    Noise predictor & 148 157\\
    \midrule
    Total & 823 617 \\
    \bottomrule
    \end{tabular*}
    \caption{Network hyperparameters of the \textsc{CaloGraph} model}
    \label{tab:hyperparams}
\end{table}
%
\section{Results}
\begin{figure}[t]
    \centering
    \begin{subfigure}{0.49\textwidth}
        \centering
        \includegraphics[page=1, width=\linewidth]{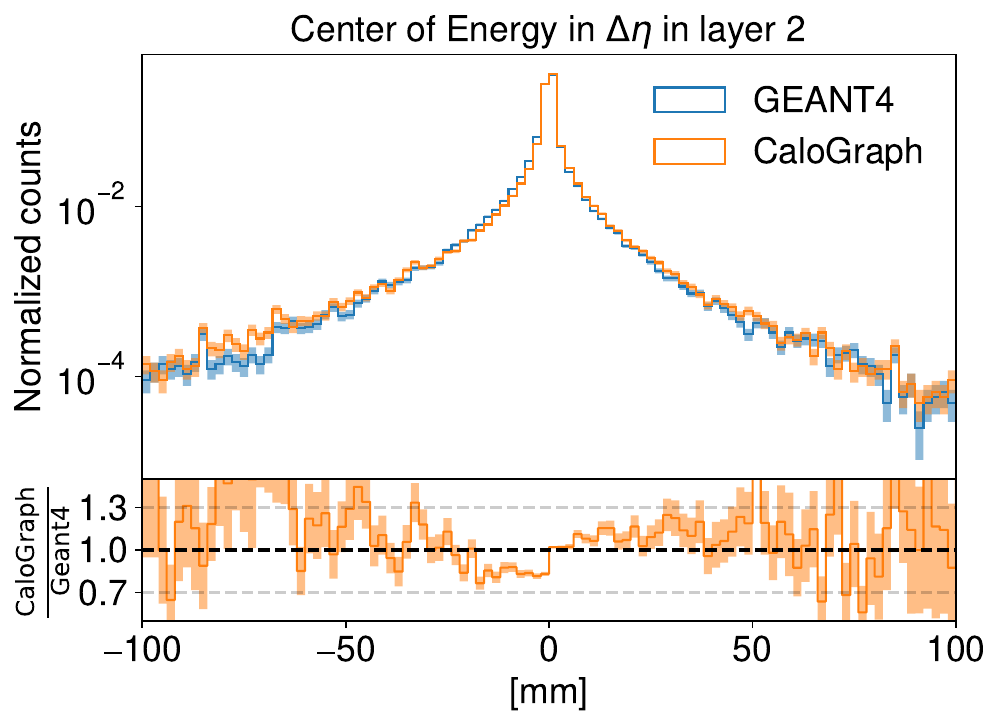} 
    \end{subfigure}
    \hfill
    \begin{subfigure}{0.49\textwidth}
        \centering
        \includegraphics[page=2, width=\linewidth]{figs/hlf_phot.pdf} 
    \end{subfigure}
    \begin{subfigure}{0.49\textwidth}
        \centering
        \includegraphics[page=3, width=\linewidth]{figs/hlf_phot.pdf} 
    \end{subfigure}
    \hspace{1pt}
    \begin{subfigure}{0.49\textwidth}
        \centering
        \includegraphics[page=4, width=\linewidth]{figs/hlf_phot.pdf} 
    \end{subfigure}
    \caption{Distribution high-level features in the photons sample.}
    \label{fig:distr-phot}
\end{figure}

\begin{figure}[t]
    \centering
    \begin{subfigure}{0.49\textwidth}
        \centering
        \includegraphics[page=1, width=\linewidth]{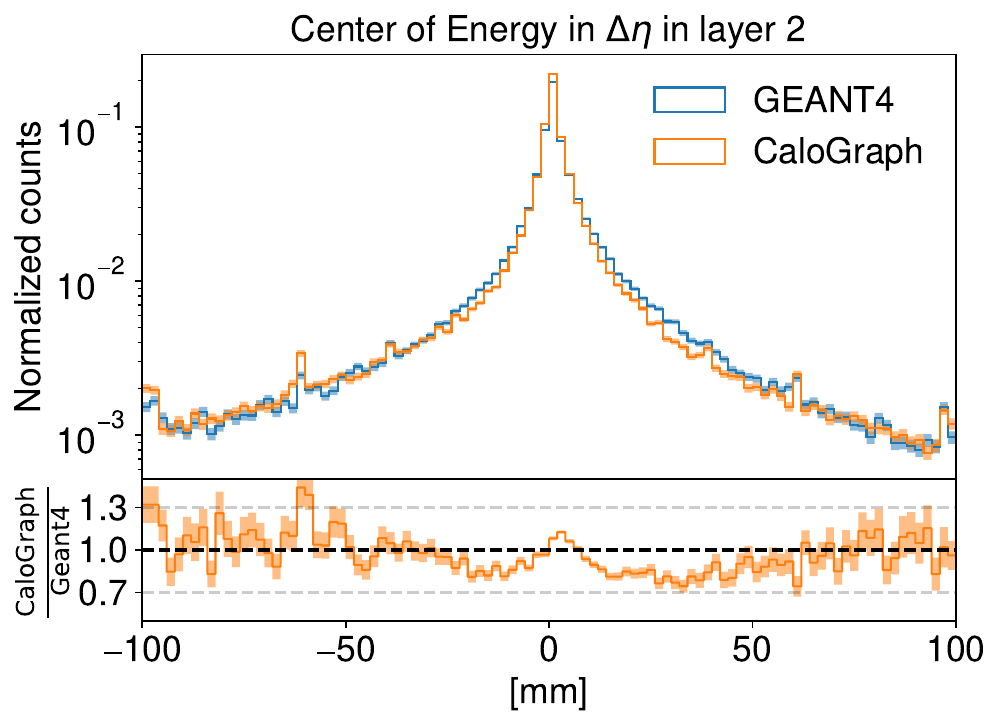} 
    \end{subfigure}
    \hfill
    \begin{subfigure}{0.49\textwidth}
        \centering
        \includegraphics[page=2, width=\linewidth]{figs/hlf_pion.pdf} 
    \end{subfigure}
    \begin{subfigure}{0.49\textwidth}
        \centering
        \includegraphics[page=3, width=\linewidth]{figs/hlf_pion.pdf} 
    \end{subfigure}
    \hspace{1pt}
    \begin{subfigure}{0.49\textwidth}
        \centering
        \includegraphics[page=4, width=\linewidth]{figs/hlf_pion.pdf} 
    \end{subfigure}
    \caption{Distribution high-level features in the pions sample.}
    \label{fig:distr-pion}
\end{figure}

To evaluate the performance of \textsc{CaloGraph}, we consider different high-level features that reflect how well the shower shapes are learned. We can define the centre of energy and its width for a layer $n$ in the angular direction $\eta$:
\begin{equation}
    \langle \eta \rangle_n = \frac{\sum_i\eta_{n,i}E_{n,i}}{\sum_iE_{n,i}},\qquad\qquad \sigma_{\langle \eta \rangle,n}=\sqrt{\frac{\sum_i\eta_{n,i}^2E_{n,i}}{\sum_iE_{n,i}}-\langle \eta \rangle^2_n}
\label{eq:highfeat}
\end{equation}
where the sum over $i$ goes over all layer voxels and $E_i$ is the generated energy deposition of voxel $i$. We show these quantities for layer 2 in the upper panels of Fig.~\ref{fig:distr-phot} for the photon test dataset and respectively Fig.~\ref{fig:distr-pion} for the pion test dataset. The distributions are modelled well for both datasets, with a percent-level discrepancy with respect to the \textsc{Geant4} baseline. The peak at 0 width stems from events that have at most one active cell in the layer.

\begin{figure}[t]
    \centering
    \includegraphics[width=\textwidth]{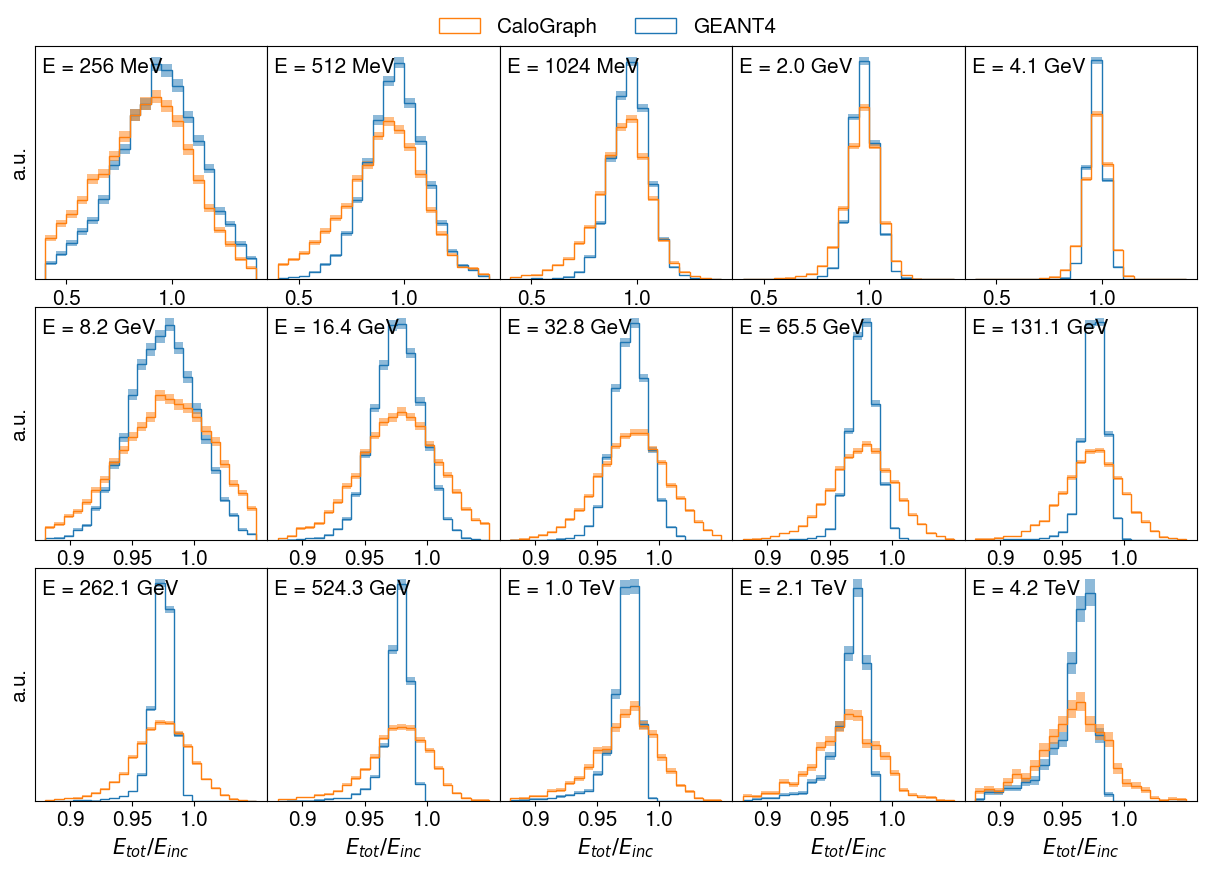}
    \caption{Ratio of total energy to the discrete values of $E_{inc}$ in the photons sample.}
    \label{fig:etot-einc-phot}
\end{figure}

\begin{figure}[t]
    \centering
    \includegraphics[width=\textwidth]{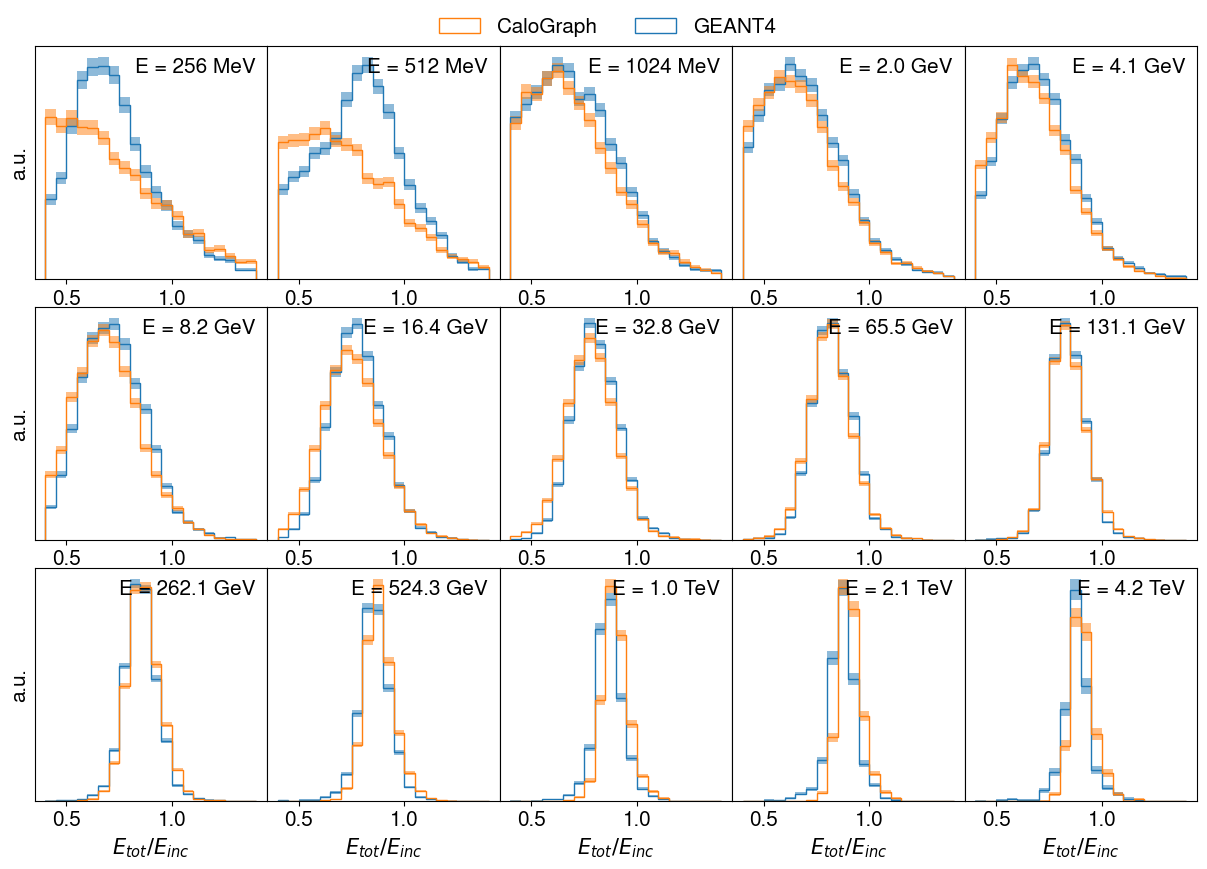}
    \caption{Ratio of total energy to the discrete values of $E_{inc}$ in the pions sample.}
    \label{fig:etot-einc-pion}
\end{figure}

The lower panels give insights into the energy modelling. On the left, we see the energy distribution in layer 2. For the photon sample, we see steps in the histogram at high energies, which are an artefact of the discrete incoming energies in the dataset. Apart from the discrepancy in some of those bins, the network learns to reproduce the correct distribution. The same can be said for the pion dataset, except for the mismodelling at energies below 500 MeV. A shortcoming of the network can be observed in the total energy modelling. For the photon dataset, the \textsc{Geant4} distribution of the total deposited energy normalized over the incoming energy is sharply peaked near 1. The network prediction is similarly peaked but creates a wider distribution. This can be seen in more detail in Fig.~\ref{fig:etot-einc-phot}, where the ratio is shown for different incident energies. The modelling is better for low incident energies where the distributions are wider. The performance on the pion dataset is better. Due to non-compensation and larger energy fluctuations associated with hadronic showers, the total energy distribution is wider than for photons and is captured better by the network. Overall, we have, at most, a discrepancy of 15\% in the bulk of the data. The distributions for separated incident energies in Fig.~\ref{fig:etot-einc-pion} are well-modelled for all energies above 500 MeV, the energy region that appears to be problematic also for the per-layer distributions. 

An additional useful metric for the evaluation of a generative model is the performance of a binary classifier trained to distinguish between \textsc{Geant4} and generated samples. The better the generative model, the harder it is to distinguish them, and the worse the performance of the classifier. We train two different classifiers for each dataset, one using low-level features and the other using high-level features. For both cases, the network is a simple MLP with two hidden layers with a size of 512. The low-level classifier takes as input only the incoming energy and the voxel energies as a list. For the high-level features, the network focuses more on the shower shape variables than on the energy values only. We combine the centres of energy and their widths in $\eta$ and $\phi$, which are defined in analogy to Eq.~\ref{eq:highfeat}, together with the deposited energies in each layer and the incoming energy and pass it as input to the classifier. The classifiers are trained on a sample containing equal proportions of showers from \textsc{CaloGraph} and \textsc{Geant4}. The area under the receiver operating characteristic curve (AUC) for the different classifiers for the pion and photon dataset are displayed in Tab.~\ref{tab:auc-scores}. A perfect generative model would result in an AUC of 0.5. For both datasets, the high-level classifier shows better results than the low-level one, reflecting the previously discussed observation that the network correctly learns the shower shape, whereas the total energy prediction shows room for improvement. 
We can understand that, in this case, the AUC on the photon dataset is lower than for the pions since the lower number of voxels makes it easier to learn the shape. We find that the low-level classifiers' performance is comparable for both datasets.
\begin{table}[b]
    \centering
    \begin{tabular}{lcc}
         \toprule
         Dataset & Low-level features & High-level features\\
         \midrule
         photons & 0.81 & 0.63 \\
         pions & 0.80 & 0.72\\
         \bottomrule
    \end{tabular}
    \caption{The AUC values for the DNN classifier trained to distinguish GEANT4 and generated showers for both datasets.}
    \label{tab:auc-scores}
\end{table}
%
%

In addition to the accuracy of shower generation, it is critical to look at the timing. In the tstacarch for fast surrogates, we want to minimize the generation time per shower as much as possible. In general, while diffusion models exceed competitive approaches in performance, their disadvantage is the slow generation time caused by the iterative process of noise removal. This can be mitigated with innovative sampling methods~\cite{chooseyourdiffusion}, for example, progressive distillation as in~\cite{caloscorev2}. The method involves training an additional model that learns to halve the steps each time, which significantly speeds up the generation but features a loss in performance. For \textsc{CaloGraph}, however, further speedup does not seem necessary. The generation times per shower for both datasets are depicted in Tab.~\ref{tab:gen-time} for different batch sizes on CPU and GPU. We attribute our low inference time to the efficient graph representation of the data and the small size of the network compared to other diffusion models.
\begin{table}[b]
    \centering
    \begin{tabular}{l|cc|cc}
         \toprule
         & \multicolumn{2}{@{}c|}{photons} & \multicolumn{2}{@{}c}{pions} \\
           Batch size & GPU & CPU & GPU & CPU\\
         \midrule
         1 & 1.57 & 1.92 & 1.52 & 2.07\\
         10 & 0.17 & 0.30 & 0.17 & 0.37\\
         100 & 0.03 & 0.12 & 0.03 & 0.20\\
         1000 & 0.01 & 0.17 & 0.02 & 0.35\\
         \bottomrule
    \end{tabular}
    \caption{Generation time per shower in seconds for different batch sizes for both datasets. The GPU studies were run on an NVIDIA TITAN RTX with 24GB VRAM, and for the CPU on an Intel(R) Xeon(R) Gold 6234 at 3.30GHz}
    \label{tab:gen-time}
\end{table}

\section{Conclusion}
We presented a novel diffusion model utilizing a graph neural network for the task of generating calorimeter showers and demonstrated the performance on the photon and pion dataset 1 of the \fastcal. In the case of irregular cell geometry, representing the calorimeter as a graph simplifies the pre- and postprocessing of the input data significantly and allows the generation of correct shower shapes due to the graph connectivity and the passage of information between neighbouring cells. Despite the use of diffusion, the inference time is fast due to the compactness of the model. A shortcoming of the approach is the slight mismodeling of the total deposited energy, especially in cases where it follows a sharp distribution, as in the photon dataset. This can potentially be mitigated in future work by adding a network that predicts the total energy per layer before diffusing it among the voxels. We also note that with the graph structure connecting all nearest neighbours, this model can primarily be used for low-granularity detectors only. Considering different connectivity, significantly reducing the number of edges and, therefore, also the memory usage might allow a generalization of the method. 

\section*{Acknowledgement}
We are grateful for the support provided by the collaborative Weizmann Institute and MBZUAI research grant, as well as the Benoziyo Center for High Energy Physics. Special thanks go to our colleagues at MBZUAI, especially Shangsong Liang and Ruihong Zeng, for their insightful discussions. We also extend our gratitude to Edward Shields for his valuable contribution.
\clearpage
\bibliography{fastsim}

\begin{thebibliography}{46}%
\makeatletter
\providecommand \@ifxundefined [1]{%
 \@ifx{#1\undefined}
}%
\providecommand \@ifnum [1]{%
 \ifnum #1\expandafter \@firstoftwo
 \else \expandafter \@secondoftwo
 \fi
}%
\providecommand \@ifx [1]{%
 \ifx #1\expandafter \@firstoftwo
 \else \expandafter \@secondoftwo
 \fi
}%
\providecommand \natexlab [1]{#1}%
\providecommand \enquote  [1]{``#1''}%
\providecommand \bibnamefont  [1]{#1}%
\providecommand \bibfnamefont [1]{#1}%
\providecommand \citenamefont [1]{#1}%
\providecommand \href@noop [0]{\@secondoftwo}%
\providecommand \href [0]{\begingroup \@sanitize@url \@href}%
\providecommand \@href[1]{\@@startlink{#1}\@@href}%
\providecommand \@@href[1]{\endgroup#1\@@endlink}%
\providecommand \@sanitize@url [0]{\catcode `\\12\catcode `\$12\catcode `\&12\catcode `\#12\catcode `\^12\catcode `\_12\catcode `\%12\relax}%
\providecommand \@@startlink[1]{}%
\providecommand \@@endlink[0]{}%
\providecommand \url  [0]{\begingroup\@sanitize@url \@url }%
\providecommand \@url [1]{\endgroup\@href {#1}{\urlprefix }}%
\providecommand \urlprefix  [0]{URL }%
\providecommand \Eprint [0]{\href }%
\providecommand \doibase [0]{https://doi.org/}%
\providecommand \selectlanguage [0]{\@gobble}%
\providecommand \bibinfo  [0]{\@secondoftwo}%
\providecommand \bibfield  [0]{\@secondoftwo}%
\providecommand \translation [1]{[#1]}%
\providecommand \BibitemOpen [0]{}%
\providecommand \bibitemStop [0]{}%
\providecommand \bibitemNoStop [0]{.\EOS\space}%
\providecommand \EOS [0]{\spacefactor3000\relax}%
\providecommand \BibitemShut  [1]{\csname bibitem#1\endcsname}%
\let\auto@bib@innerbib\@empty
\bibitem [{\citenamefont {Butter}\ \emph {et~al.}(2023)\citenamefont {Butter}, \citenamefont {Plehn}, \citenamefont {Schumann}, \citenamefont {Badger}, \citenamefont {Caron} \emph {et~al.}}]{mlandlhcgen}%
  \BibitemOpen
  \bibfield  {author} {\bibinfo {author} {\bibfnamefont {A.}~\bibnamefont {Butter}}, \bibinfo {author} {\bibfnamefont {T.}~\bibnamefont {Plehn}}, \bibinfo {author} {\bibfnamefont {S.}~\bibnamefont {Schumann}}, \bibinfo {author} {\bibfnamefont {S.}~\bibnamefont {Badger}}, \bibinfo {author} {\bibfnamefont {S.}~\bibnamefont {Caron}}, \emph {et~al.},\ }\bibfield  {title} {\bibinfo {title} {{Machine learning and LHC event generation}},\ }\href {https://doi.org/10.21468/SciPostPhys.14.4.079} {\bibfield  {journal} {\bibinfo  {journal} {SciPost Phys.}\ }\textbf {\bibinfo {volume} {14}},\ \bibinfo {pages} {079} (\bibinfo {year} {2023})}\BibitemShut {NoStop}%
\bibitem [{\citenamefont {Hashemi}\ and\ \citenamefont {Krause}(2023)}]{deepgendet}%
  \BibitemOpen
  \bibfield  {author} {\bibinfo {author} {\bibfnamefont {H.}~\bibnamefont {Hashemi}}\ and\ \bibinfo {author} {\bibfnamefont {C.}~\bibnamefont {Krause}},\ }\bibfield  {title} {\bibinfo {title} {{Deep Generative Models for Detector Signature Simulation: An Analytical Taxonomy}},\ }\href@noop {} {\  (\bibinfo {year} {2023})},\ \Eprint {https://arxiv.org/abs/2312.09597} {arXiv:2312.09597 [physics.ins-det]} \BibitemShut {NoStop}%
\bibitem [{\citenamefont {Agostinelli}\ \emph {et~al.}(2003)\citenamefont {Agostinelli} \emph {et~al.}}]{geant}%
  \BibitemOpen
  \bibfield  {author} {\bibinfo {author} {\bibfnamefont {S.}~\bibnamefont {Agostinelli}} \emph {et~al.} (\bibinfo {collaboration} {GEANT4}),\ }\bibfield  {title} {\bibinfo {title} {{GEANT4: A simulation toolkit}},\ }\href {https://doi.org/10.1016/S0168-9002(03)01368-8} {\bibfield  {journal} {\bibinfo  {journal} {Nucl. Instrum. Meth.}\ }\textbf {\bibinfo {volume} {A506}},\ \bibinfo {pages} {250} (\bibinfo {year} {2003})}\BibitemShut {NoStop}%
\bibitem [{\citenamefont {{ATLAS Collaboration}}(2010)}]{SOFT-2010-01}%
  \BibitemOpen
  \bibfield  {author} {\bibinfo {author} {\bibnamefont {{ATLAS Collaboration}}},\ }\bibfield  {title} {\bibinfo {title} {{The ATLAS Simulation Infrastructure}},\ }\href {https://doi.org/10.1140/epjc/s10052-010-1429-9} {\bibfield  {journal} {\bibinfo  {journal} {Eur. Phys. J. C}\ }\textbf {\bibinfo {volume} {70}},\ \bibinfo {pages} {823} (\bibinfo {year} {2010})}\BibitemShut {NoStop}%
\bibitem [{\citenamefont {Br\"uning}\ \emph {et~al.}(2022)\citenamefont {Br\"uning}, \citenamefont {Gray}, \citenamefont {Klein}, \citenamefont {Lamont}, \citenamefont {Narain}, \citenamefont {Polifka},\ and\ \citenamefont {Rossi}}]{hl-lhc}%
  \BibitemOpen
  \bibfield  {author} {\bibinfo {author} {\bibfnamefont {O.}~\bibnamefont {Br\"uning}}, \bibinfo {author} {\bibfnamefont {H.}~\bibnamefont {Gray}}, \bibinfo {author} {\bibfnamefont {K.}~\bibnamefont {Klein}}, \bibinfo {author} {\bibfnamefont {M.}~\bibnamefont {Lamont}}, \bibinfo {author} {\bibfnamefont {M.}~\bibnamefont {Narain}}, \bibinfo {author} {\bibfnamefont {R.}~\bibnamefont {Polifka}},\ and\ \bibinfo {author} {\bibfnamefont {L.}~\bibnamefont {Rossi}},\ }\bibfield  {title} {\bibinfo {title} {{The scientific potential and technological challenges of the High-Luminosity Large Hadron Collider program}},\ }\href {https://doi.org/10.1088/1361-6633/ac5106} {\bibfield  {journal} {\bibinfo  {journal} {Rept. Prog. Phys.}\ }\textbf {\bibinfo {volume} {85}},\ \bibinfo {pages} {046201} (\bibinfo {year} {2022})}\BibitemShut {NoStop}%
\bibitem [{\citenamefont {Heath}(2018)}]{atlasfastcalosim}%
  \BibitemOpen
  \bibfield  {author} {\bibinfo {author} {\bibfnamefont {M.~P.}\ \bibnamefont {Heath}} (\bibinfo {collaboration} {ATLAS}),\ }\bibfield  {title} {\bibinfo {title} {{The new ATLAS Fast Calorimeter Simulation}},\ }\href {https://doi.org/10.22323/1.314.0792} {\bibfield  {journal} {\bibinfo  {journal} {PoS}\ }\textbf {\bibinfo {volume} {EPS-HEP2017}},\ \bibinfo {pages} {792} (\bibinfo {year} {2018})}\BibitemShut {NoStop}%
\bibitem [{\citenamefont {Giammanco}(2014)}]{cmsfastcalosim}%
  \BibitemOpen
  \bibfield  {author} {\bibinfo {author} {\bibfnamefont {A.}~\bibnamefont {Giammanco}},\ }\bibfield  {title} {\bibinfo {title} {{The Fast Simulation of the CMS Experiment}},\ }\href {https://doi.org/10.1088/1742-6596/513/2/022012} {\bibfield  {journal} {\bibinfo  {journal} {J. Phys. Conf. Ser.}\ }\textbf {\bibinfo {volume} {513}},\ \bibinfo {pages} {022012} (\bibinfo {year} {2014})}\BibitemShut {NoStop}%
\bibitem [{\citenamefont {Paganini}\ \emph {et~al.}(2018)\citenamefont {Paganini}, \citenamefont {de~Oliveira},\ and\ \citenamefont {Nachman}}]{calogan}%
  \BibitemOpen
  \bibfield  {author} {\bibinfo {author} {\bibfnamefont {M.}~\bibnamefont {Paganini}}, \bibinfo {author} {\bibfnamefont {L.}~\bibnamefont {de~Oliveira}},\ and\ \bibinfo {author} {\bibfnamefont {B.}~\bibnamefont {Nachman}},\ }\bibfield  {title} {\bibinfo {title} {{CaloGAN : Simulating 3D high energy particle showers in multilayer electromagnetic calorimeters with generative adversarial networks}},\ }\href {https://doi.org/10.1103/PhysRevD.97.014021} {\bibfield  {journal} {\bibinfo  {journal} {Phys. Rev. D}\ }\textbf {\bibinfo {volume} {97}},\ \bibinfo {pages} {014021} (\bibinfo {year} {2018})},\ \Eprint {https://arxiv.org/abs/1712.10321} {arXiv:1712.10321 [hep-ex]} \BibitemShut {NoStop}%
\bibitem [{\citenamefont {Aad}\ \emph {et~al.}(2022)\citenamefont {Aad} \emph {et~al.}}]{ATLAS:2021pzo}%
  \BibitemOpen
  \bibfield  {author} {\bibinfo {author} {\bibfnamefont {G.}~\bibnamefont {Aad}} \emph {et~al.} (\bibinfo {collaboration} {ATLAS}),\ }\bibfield  {title} {\bibinfo {title} {{AtlFast3: the next generation of fast simulation in ATLAS}},\ }\href {https://doi.org/10.1007/s41781-021-00079-7} {\bibfield  {journal} {\bibinfo  {journal} {Comput. Softw. Big Sci.}\ }\textbf {\bibinfo {volume} {6}},\ \bibinfo {pages} {7} (\bibinfo {year} {2022})},\ \Eprint {https://arxiv.org/abs/2109.02551} {arXiv:2109.02551 [hep-ex]} \BibitemShut {NoStop}%
\bibitem [{\citenamefont {Buhmann}\ \emph {et~al.}(2023{\natexlab{a}})\citenamefont {Buhmann}, \citenamefont {Diefenbacher}, \citenamefont {Eren}, \citenamefont {Gaede}, \citenamefont {Kasieczka}, \citenamefont {Korol}, \citenamefont {Korcari}, \citenamefont {Kr\"uger},\ and\ \citenamefont {McKeown}}]{caloclouds}%
  \BibitemOpen
  \bibfield  {author} {\bibinfo {author} {\bibfnamefont {E.}~\bibnamefont {Buhmann}}, \bibinfo {author} {\bibfnamefont {S.}~\bibnamefont {Diefenbacher}}, \bibinfo {author} {\bibfnamefont {E.}~\bibnamefont {Eren}}, \bibinfo {author} {\bibfnamefont {F.}~\bibnamefont {Gaede}}, \bibinfo {author} {\bibfnamefont {G.}~\bibnamefont {Kasieczka}}, \bibinfo {author} {\bibfnamefont {A.}~\bibnamefont {Korol}}, \bibinfo {author} {\bibfnamefont {W.}~\bibnamefont {Korcari}}, \bibinfo {author} {\bibfnamefont {K.}~\bibnamefont {Kr\"uger}},\ and\ \bibinfo {author} {\bibfnamefont {P.}~\bibnamefont {McKeown}},\ }\bibfield  {title} {\bibinfo {title} {{CaloClouds: fast geometry-independent highly-granular calorimeter simulation}},\ }\href {https://doi.org/10.1088/1748-0221/18/11/P11025} {\bibfield  {journal} {\bibinfo  {journal} {JINST}\ }\textbf {\bibinfo {volume} {18}}\bibfield  {number} {\bibinfo  {number} { (11)},\ \bibinfo {pages} {P11025}},\ }\Eprint {https://arxiv.org/abs/2305.04847} {arXiv:2305.04847 [physics.ins-det]}
  \BibitemShut {NoStop}%
\bibitem [{\citenamefont {Buhmann}\ \emph {et~al.}(2023{\natexlab{b}})\citenamefont {Buhmann}, \citenamefont {Gaede}, \citenamefont {Kasieczka}, \citenamefont {Korol}, \citenamefont {Korcari}, \citenamefont {Kr\"uger},\ and\ \citenamefont {McKeown}}]{caloclouds2}%
  \BibitemOpen
  \bibfield  {author} {\bibinfo {author} {\bibfnamefont {E.}~\bibnamefont {Buhmann}}, \bibinfo {author} {\bibfnamefont {F.}~\bibnamefont {Gaede}}, \bibinfo {author} {\bibfnamefont {G.}~\bibnamefont {Kasieczka}}, \bibinfo {author} {\bibfnamefont {A.}~\bibnamefont {Korol}}, \bibinfo {author} {\bibfnamefont {W.}~\bibnamefont {Korcari}}, \bibinfo {author} {\bibfnamefont {K.}~\bibnamefont {Kr\"uger}},\ and\ \bibinfo {author} {\bibfnamefont {P.}~\bibnamefont {McKeown}},\ }\bibfield  {title} {\bibinfo {title} {{CaloClouds II: Ultra-Fast Geometry-Independent Highly-Granular Calorimeter Simulation}},\ }\href@noop {} {\  (\bibinfo {year} {2023}{\natexlab{b}})},\ \Eprint {https://arxiv.org/abs/2309.05704} {arXiv:2309.05704 [physics.ins-det]} \BibitemShut {NoStop}%
\bibitem [{\citenamefont {Amram}\ and\ \citenamefont {Pedro}(2023)}]{calodiffusion}%
  \BibitemOpen
  \bibfield  {author} {\bibinfo {author} {\bibfnamefont {O.}~\bibnamefont {Amram}}\ and\ \bibinfo {author} {\bibfnamefont {K.}~\bibnamefont {Pedro}},\ }\bibfield  {title} {\bibinfo {title} {{CaloDiffusion with GLaM for High Fidelity Calorimeter Simulation}},\ }\href@noop {} {\  (\bibinfo {year} {2023})},\ \Eprint {https://arxiv.org/abs/2308.03876} {arXiv:2308.03876 [physics.ins-det]} \BibitemShut {NoStop}%
\bibitem [{\citenamefont {Krause}\ \emph {et~al.}(2022)\citenamefont {Krause}, \citenamefont {Pang},\ and\ \citenamefont {Shih}}]{caloflow_ds1}%
  \BibitemOpen
  \bibfield  {author} {\bibinfo {author} {\bibfnamefont {C.}~\bibnamefont {Krause}}, \bibinfo {author} {\bibfnamefont {I.}~\bibnamefont {Pang}},\ and\ \bibinfo {author} {\bibfnamefont {D.}~\bibnamefont {Shih}},\ }\bibfield  {title} {\bibinfo {title} {{CaloFlow for CaloChallenge Dataset 1}},\ }\href@noop {} {\  (\bibinfo {year} {2022})},\ \Eprint {https://arxiv.org/abs/2210.14245} {arXiv:2210.14245 [physics.ins-det]} \BibitemShut {NoStop}%
\bibitem [{\citenamefont {Ernst}\ \emph {et~al.}(2023)\citenamefont {Ernst}, \citenamefont {Favaro}, \citenamefont {Krause}, \citenamefont {Plehn},\ and\ \citenamefont {Shih}}]{caloinn}%
  \BibitemOpen
  \bibfield  {author} {\bibinfo {author} {\bibfnamefont {F.}~\bibnamefont {Ernst}}, \bibinfo {author} {\bibfnamefont {L.}~\bibnamefont {Favaro}}, \bibinfo {author} {\bibfnamefont {C.}~\bibnamefont {Krause}}, \bibinfo {author} {\bibfnamefont {T.}~\bibnamefont {Plehn}},\ and\ \bibinfo {author} {\bibfnamefont {D.}~\bibnamefont {Shih}},\ }\bibfield  {title} {\bibinfo {title} {{Normalizing Flows for High-Dimensional Detector Simulations}},\ }\href@noop {} {\  (\bibinfo {year} {2023})},\ \Eprint {https://arxiv.org/abs/2312.09290} {arXiv:2312.09290 [hep-ph]} \BibitemShut {NoStop}%
\bibitem [{\citenamefont {Mikuni}\ and\ \citenamefont {Nachman}(2022)}]{caloscore}%
  \BibitemOpen
  \bibfield  {author} {\bibinfo {author} {\bibfnamefont {V.}~\bibnamefont {Mikuni}}\ and\ \bibinfo {author} {\bibfnamefont {B.}~\bibnamefont {Nachman}},\ }\bibfield  {title} {\bibinfo {title} {{Score-based generative models for calorimeter shower simulation}},\ }\href {https://doi.org/10.1103/PhysRevD.106.092009} {\bibfield  {journal} {\bibinfo  {journal} {Phys. Rev. D}\ }\textbf {\bibinfo {volume} {106}},\ \bibinfo {pages} {092009} (\bibinfo {year} {2022})},\ \Eprint {https://arxiv.org/abs/2206.11898} {arXiv:2206.11898 [hep-ph]} \BibitemShut {NoStop}%
\bibitem [{\citenamefont {Mikuni}\ and\ \citenamefont {Nachman}(2023)}]{caloscorev2}%
  \BibitemOpen
  \bibfield  {author} {\bibinfo {author} {\bibfnamefont {V.}~\bibnamefont {Mikuni}}\ and\ \bibinfo {author} {\bibfnamefont {B.}~\bibnamefont {Nachman}},\ }\bibfield  {title} {\bibinfo {title} {{CaloScore v2: Single-shot Calorimeter Shower Simulation with Diffusion Models}},\ }\href@noop {} {\  (\bibinfo {year} {2023})},\ \Eprint {https://arxiv.org/abs/2308.03847} {arXiv:2308.03847 [hep-ph]} \BibitemShut {NoStop}%
\bibitem [{\citenamefont {Faucci~Giannelli}\ and\ \citenamefont {Zhang}(2023)}]{caloshowergan}%
  \BibitemOpen
  \bibfield  {author} {\bibinfo {author} {\bibfnamefont {M.}~\bibnamefont {Faucci~Giannelli}}\ and\ \bibinfo {author} {\bibfnamefont {R.}~\bibnamefont {Zhang}},\ }\bibfield  {title} {\bibinfo {title} {{CaloShowerGAN, a Generative Adversarial Networks model for fast calorimeter shower simulation}},\ }\href@noop {} {\  (\bibinfo {year} {2023})},\ \Eprint {https://arxiv.org/abs/2309.06515} {arXiv:2309.06515 [physics.ins-det]} \BibitemShut {NoStop}%
\bibitem [{\citenamefont {Pang}\ \emph {et~al.}(2023)\citenamefont {Pang}, \citenamefont {Raine},\ and\ \citenamefont {Shih}}]{supercalo}%
  \BibitemOpen
  \bibfield  {author} {\bibinfo {author} {\bibfnamefont {I.}~\bibnamefont {Pang}}, \bibinfo {author} {\bibfnamefont {J.~A.}\ \bibnamefont {Raine}},\ and\ \bibinfo {author} {\bibfnamefont {D.}~\bibnamefont {Shih}},\ }\bibfield  {title} {\bibinfo {title} {{SuperCalo: Calorimeter shower super-resolution}},\ }\href@noop {} {\  (\bibinfo {year} {2023})},\ \Eprint {https://arxiv.org/abs/2308.11700} {arXiv:2308.11700 [physics.ins-det]} \BibitemShut {NoStop}%
\bibitem [{\citenamefont {Buckley}\ \emph {et~al.}(2023)\citenamefont {Buckley}, \citenamefont {Krause}, \citenamefont {Pang},\ and\ \citenamefont {Shih}}]{inductive_caloflow}%
  \BibitemOpen
  \bibfield  {author} {\bibinfo {author} {\bibfnamefont {M.~R.}\ \bibnamefont {Buckley}}, \bibinfo {author} {\bibfnamefont {C.}~\bibnamefont {Krause}}, \bibinfo {author} {\bibfnamefont {I.}~\bibnamefont {Pang}},\ and\ \bibinfo {author} {\bibfnamefont {D.}~\bibnamefont {Shih}},\ }\bibfield  {title} {\bibinfo {title} {{Inductive CaloFlow}},\ }\href@noop {} {\  (\bibinfo {year} {2023})},\ \Eprint {https://arxiv.org/abs/2305.11934} {arXiv:2305.11934 [physics.ins-det]} \BibitemShut {NoStop}%
\bibitem [{\citenamefont {Giannelli}\ \emph {et~al.}(2022)\citenamefont {Giannelli}, \citenamefont {Kasieczka}, \citenamefont {Krause}, \citenamefont {Nachman}, \citenamefont {Salamani}, \citenamefont {Shih},\ and\ \citenamefont {Zaborowska}}]{ds1}%
  \BibitemOpen
  \bibfield  {author} {\bibinfo {author} {\bibfnamefont {M.~F.}\ \bibnamefont {Giannelli}}, \bibinfo {author} {\bibfnamefont {G.}~\bibnamefont {Kasieczka}}, \bibinfo {author} {\bibfnamefont {C.}~\bibnamefont {Krause}}, \bibinfo {author} {\bibfnamefont {B.}~\bibnamefont {Nachman}}, \bibinfo {author} {\bibfnamefont {D.}~\bibnamefont {Salamani}}, \bibinfo {author} {\bibfnamefont {D.}~\bibnamefont {Shih}},\ and\ \bibinfo {author} {\bibfnamefont {A.}~\bibnamefont {Zaborowska}},\ }\bibfield  {title} {\bibinfo {title} {{Fast Calorimeter Simulation Challenge 2022 - Dataset 1}},\ }\href {https://doi.org/10.5281/zenodo.6368338} {10.5281/zenodo.6368338} (\bibinfo {year} {2022})\BibitemShut {NoStop}%
\bibitem [{\citenamefont {Faucci~Giannelli}\ \emph {et~al.}(2022{\natexlab{a}})\citenamefont {Faucci~Giannelli}, \citenamefont {Kasieczka}, \citenamefont {Krause}, \citenamefont {Nachman}, \citenamefont {Salamani}, \citenamefont {Shih},\ and\ \citenamefont {Zaborowska}}]{ds2}%
  \BibitemOpen
  \bibfield  {author} {\bibinfo {author} {\bibfnamefont {M.}~\bibnamefont {Faucci~Giannelli}}, \bibinfo {author} {\bibfnamefont {G.}~\bibnamefont {Kasieczka}}, \bibinfo {author} {\bibfnamefont {C.}~\bibnamefont {Krause}}, \bibinfo {author} {\bibfnamefont {B.}~\bibnamefont {Nachman}}, \bibinfo {author} {\bibfnamefont {D.}~\bibnamefont {Salamani}}, \bibinfo {author} {\bibfnamefont {D.}~\bibnamefont {Shih}},\ and\ \bibinfo {author} {\bibfnamefont {A.}~\bibnamefont {Zaborowska}},\ }\bibfield  {title} {\bibinfo {title} {{Fast Calorimeter Simulation Challenge 2022 - Dataset 2}},\ }\href {https://doi.org/10.5281/zenodo.6366271} {10.5281/zenodo.6366271} (\bibinfo {year} {2022}{\natexlab{a}})\BibitemShut {NoStop}%
\bibitem [{\citenamefont {Faucci~Giannelli}\ \emph {et~al.}(2022{\natexlab{b}})\citenamefont {Faucci~Giannelli}, \citenamefont {Kasieczka}, \citenamefont {Krause}, \citenamefont {Nachman}, \citenamefont {Salamani}, \citenamefont {Shih},\ and\ \citenamefont {Zaborowska}}]{ds3}%
  \BibitemOpen
  \bibfield  {author} {\bibinfo {author} {\bibfnamefont {M.}~\bibnamefont {Faucci~Giannelli}}, \bibinfo {author} {\bibfnamefont {G.}~\bibnamefont {Kasieczka}}, \bibinfo {author} {\bibfnamefont {C.}~\bibnamefont {Krause}}, \bibinfo {author} {\bibfnamefont {B.}~\bibnamefont {Nachman}}, \bibinfo {author} {\bibfnamefont {D.}~\bibnamefont {Salamani}}, \bibinfo {author} {\bibfnamefont {D.}~\bibnamefont {Shih}},\ and\ \bibinfo {author} {\bibfnamefont {A.}~\bibnamefont {Zaborowska}},\ }\bibfield  {title} {\bibinfo {title} {{Fast Calorimeter Simulation Challenge 2022 - Dataset 3}},\ }\href {https://doi.org/10.5281/zenodo.6366324} {10.5281/zenodo.6366324} (\bibinfo {year} {2022}{\natexlab{b}})\BibitemShut {NoStop}%
\bibitem [{\citenamefont {Mikuni}\ \emph {et~al.}(2023)\citenamefont {Mikuni}, \citenamefont {Nachman},\ and\ \citenamefont {Pettee}}]{jetben}%
  \BibitemOpen
  \bibfield  {author} {\bibinfo {author} {\bibfnamefont {V.}~\bibnamefont {Mikuni}}, \bibinfo {author} {\bibfnamefont {B.}~\bibnamefont {Nachman}},\ and\ \bibinfo {author} {\bibfnamefont {M.}~\bibnamefont {Pettee}},\ }\bibfield  {title} {\bibinfo {title} {{Fast point cloud generation with diffusion models in high energy physics}},\ }\href {https://doi.org/10.1103/PhysRevD.108.036025} {\bibfield  {journal} {\bibinfo  {journal} {Phys. Rev. D}\ }\textbf {\bibinfo {volume} {108}},\ \bibinfo {pages} {036025} (\bibinfo {year} {2023})},\ \Eprint {https://arxiv.org/abs/2304.01266} {arXiv:2304.01266 [hep-ph]} \BibitemShut {NoStop}%
\bibitem [{\citenamefont {Buhmann}\ \emph {et~al.}(2023{\natexlab{c}})\citenamefont {Buhmann}, \citenamefont {Ewen}, \citenamefont {Faroughy}, \citenamefont {Golling}, \citenamefont {Kasieczka}, \citenamefont {Leigh}, \citenamefont {Qu\'etant}, \citenamefont {Raine}, \citenamefont {Sengupta},\ and\ \citenamefont {Shih}}]{jetepic}%
  \BibitemOpen
  \bibfield  {author} {\bibinfo {author} {\bibfnamefont {E.}~\bibnamefont {Buhmann}}, \bibinfo {author} {\bibfnamefont {C.}~\bibnamefont {Ewen}}, \bibinfo {author} {\bibfnamefont {D.~A.}\ \bibnamefont {Faroughy}}, \bibinfo {author} {\bibfnamefont {T.}~\bibnamefont {Golling}}, \bibinfo {author} {\bibfnamefont {G.}~\bibnamefont {Kasieczka}}, \bibinfo {author} {\bibfnamefont {M.}~\bibnamefont {Leigh}}, \bibinfo {author} {\bibfnamefont {G.}~\bibnamefont {Qu\'etant}}, \bibinfo {author} {\bibfnamefont {J.~A.}\ \bibnamefont {Raine}}, \bibinfo {author} {\bibfnamefont {D.}~\bibnamefont {Sengupta}},\ and\ \bibinfo {author} {\bibfnamefont {D.}~\bibnamefont {Shih}},\ }\bibfield  {title} {\bibinfo {title} {{EPiC-ly Fast Particle Cloud Generation with Flow-Matching and Diffusion}},\ }\href@noop {} {\  (\bibinfo {year} {2023}{\natexlab{c}})},\ \Eprint {https://arxiv.org/abs/2310.00049} {arXiv:2310.00049 [hep-ph]} \BibitemShut {NoStop}%
\bibitem [{\citenamefont {Buhmann}\ \emph {et~al.}(2023{\natexlab{d}})\citenamefont {Buhmann}, \citenamefont {Kasieczka},\ and\ \citenamefont {Thaler}}]{epicGAN}%
  \BibitemOpen
  \bibfield  {author} {\bibinfo {author} {\bibfnamefont {E.}~\bibnamefont {Buhmann}}, \bibinfo {author} {\bibfnamefont {G.}~\bibnamefont {Kasieczka}},\ and\ \bibinfo {author} {\bibfnamefont {J.}~\bibnamefont {Thaler}},\ }\bibfield  {title} {\bibinfo {title} {{EPiC-GAN: Equivariant Point Cloud Generation for Particle Jets}},\ }\href@noop {} {\  (\bibinfo {year} {2023}{\natexlab{d}})},\ \Eprint {https://arxiv.org/abs/2301.08128} {arXiv:2301.08128 [hep-ph]} \BibitemShut {NoStop}%
\bibitem [{\citenamefont {Leigh}\ \emph {et~al.}(2023)\citenamefont {Leigh}, \citenamefont {Sengupta}, \citenamefont {Raine}, \citenamefont {Qu\'etant},\ and\ \citenamefont {Golling}}]{pcdroid}%
  \BibitemOpen
  \bibfield  {author} {\bibinfo {author} {\bibfnamefont {M.}~\bibnamefont {Leigh}}, \bibinfo {author} {\bibfnamefont {D.}~\bibnamefont {Sengupta}}, \bibinfo {author} {\bibfnamefont {J.~A.}\ \bibnamefont {Raine}}, \bibinfo {author} {\bibfnamefont {G.}~\bibnamefont {Qu\'etant}},\ and\ \bibinfo {author} {\bibfnamefont {T.}~\bibnamefont {Golling}},\ }\bibfield  {title} {\bibinfo {title} {{PC-Droid: Faster diffusion and improved quality for particle cloud generation}},\ }\href@noop {} {\  (\bibinfo {year} {2023})},\ \Eprint {https://arxiv.org/abs/2307.06836} {arXiv:2307.06836 [hep-ex]} \BibitemShut {NoStop}%
\bibitem [{\citenamefont {Käch}\ \emph {et~al.}(2022)\citenamefont {Käch}, \citenamefont {Krücker},\ and\ \citenamefont {Melzer-Pellmann}}]{pointcloudtrfo}%
  \BibitemOpen
  \bibfield  {author} {\bibinfo {author} {\bibfnamefont {B.}~\bibnamefont {Käch}}, \bibinfo {author} {\bibfnamefont {D.}~\bibnamefont {Krücker}},\ and\ \bibinfo {author} {\bibfnamefont {I.}~\bibnamefont {Melzer-Pellmann}},\ }\href@noop {} {\bibinfo {title} {Point cloud generation using transformer encoders and normalising flows}} (\bibinfo {year} {2022}),\ \Eprint {https://arxiv.org/abs/2211.13623} {arXiv:2211.13623 [hep-ex]} \BibitemShut {NoStop}%
\bibitem [{\citenamefont {Simon}(2012)}]{clic}%
  \BibitemOpen
  \bibfield  {author} {\bibinfo {author} {\bibfnamefont {F.}~\bibnamefont {Simon}},\ }\bibfield  {title} {\bibinfo {title} {{Detector Systems at CLIC}},\ }\href {https://doi.org/10.1016/j.phpro.2012.02.357} {\bibfield  {journal} {\bibinfo  {journal} {Phys. Procedia}\ }\textbf {\bibinfo {volume} {37}},\ \bibinfo {pages} {63} (\bibinfo {year} {2012})},\ \Eprint {https://arxiv.org/abs/1109.3387} {arXiv:1109.3387 [physics.ins-det]} \BibitemShut {NoStop}%
\bibitem [{\citenamefont {Shlomi}\ \emph {et~al.}(2020)\citenamefont {Shlomi}, \citenamefont {Battaglia},\ and\ \citenamefont {vlimant}}]{Shlomi_2020}%
  \BibitemOpen
  \bibfield  {author} {\bibinfo {author} {\bibfnamefont {J.}~\bibnamefont {Shlomi}}, \bibinfo {author} {\bibfnamefont {P.}~\bibnamefont {Battaglia}},\ and\ \bibinfo {author} {\bibfnamefont {j.-r.}\ \bibnamefont {vlimant}},\ }\bibfield  {title} {\bibinfo {title} {Graph neural networks in particle physics},\ }\bibfield  {journal} {\bibinfo  {journal} {Machine Learning: Science and Technology}\ }\href {https://doi.org/10.1088/2632-2153/abbf9a} {10.1088/2632-2153/abbf9a} (\bibinfo {year} {2020})\BibitemShut {NoStop}%
\bibitem [{\citenamefont {DeZoort}\ \emph {et~al.}(2021)\citenamefont {DeZoort}, \citenamefont {Thais}, \citenamefont {Duarte}, \citenamefont {Razavimaleki}, \citenamefont {Atkinson}, \citenamefont {Ojalvo}, \citenamefont {Neubauer},\ and\ \citenamefont {Elmer}}]{graphtrack}%
  \BibitemOpen
  \bibfield  {author} {\bibinfo {author} {\bibfnamefont {G.}~\bibnamefont {DeZoort}}, \bibinfo {author} {\bibfnamefont {S.}~\bibnamefont {Thais}}, \bibinfo {author} {\bibfnamefont {J.}~\bibnamefont {Duarte}}, \bibinfo {author} {\bibfnamefont {V.}~\bibnamefont {Razavimaleki}}, \bibinfo {author} {\bibfnamefont {M.}~\bibnamefont {Atkinson}}, \bibinfo {author} {\bibfnamefont {I.}~\bibnamefont {Ojalvo}}, \bibinfo {author} {\bibfnamefont {M.}~\bibnamefont {Neubauer}},\ and\ \bibinfo {author} {\bibfnamefont {P.}~\bibnamefont {Elmer}},\ }\bibfield  {title} {\bibinfo {title} {Charged particle tracking via edge-classifying interaction networks},\ }\bibfield  {journal} {\bibinfo  {journal} {Computing and Software for Big Science}\ }\textbf {\bibinfo {volume} {5}},\ \href {https://doi.org/10.1007/s41781-021-00073-z} {10.1007/s41781-021-00073-z} (\bibinfo {year} {2021})\BibitemShut {NoStop}%
\bibitem [{\citenamefont {Murnane}\ \emph {et~al.}(2023)\citenamefont {Murnane}, \citenamefont {Thais},\ and\ \citenamefont {Thete}}]{graphtrack2}%
  \BibitemOpen
  \bibfield  {author} {\bibinfo {author} {\bibfnamefont {D.}~\bibnamefont {Murnane}}, \bibinfo {author} {\bibfnamefont {S.}~\bibnamefont {Thais}},\ and\ \bibinfo {author} {\bibfnamefont {A.}~\bibnamefont {Thete}},\ }\href@noop {} {\bibinfo {title} {Equivariant graph neural networks for charged particle tracking}} (\bibinfo {year} {2023}),\ \Eprint {https://arxiv.org/abs/2304.05293} {arXiv:2304.05293 [physics.ins-det]} \BibitemShut {NoStop}%
\bibitem [{\citenamefont {Liu}\ \emph {et~al.}(2023{\natexlab{a}})\citenamefont {Liu}, \citenamefont {Calafiura}, \citenamefont {Farrell}, \citenamefont {Ju}, \citenamefont {Murnane},\ and\ \citenamefont {Pham}}]{graphtrack3}%
  \BibitemOpen
  \bibfield  {author} {\bibinfo {author} {\bibfnamefont {R.}~\bibnamefont {Liu}}, \bibinfo {author} {\bibfnamefont {P.}~\bibnamefont {Calafiura}}, \bibinfo {author} {\bibfnamefont {S.}~\bibnamefont {Farrell}}, \bibinfo {author} {\bibfnamefont {X.}~\bibnamefont {Ju}}, \bibinfo {author} {\bibfnamefont {D.~T.}\ \bibnamefont {Murnane}},\ and\ \bibinfo {author} {\bibfnamefont {T.~M.}\ \bibnamefont {Pham}},\ }\href@noop {} {\bibinfo {title} {Hierarchical graph neural networks for particle track reconstruction}} (\bibinfo {year} {2023}{\natexlab{a}}),\ \Eprint {https://arxiv.org/abs/2303.01640} {arXiv:2303.01640 [hep-ex]} \BibitemShut {NoStop}%
\bibitem [{\citenamefont {Ma}\ \emph {et~al.}(2023)\citenamefont {Ma}, \citenamefont {Liu},\ and\ \citenamefont {Li}}]{jettag}%
  \BibitemOpen
  \bibfield  {author} {\bibinfo {author} {\bibfnamefont {F.}~\bibnamefont {Ma}}, \bibinfo {author} {\bibfnamefont {F.}~\bibnamefont {Liu}},\ and\ \bibinfo {author} {\bibfnamefont {W.}~\bibnamefont {Li}},\ }\bibfield  {title} {\bibinfo {title} {Jet tagging algorithm of graph network with haar pooling message passing},\ }\bibfield  {journal} {\bibinfo  {journal} {Physical Review D}\ }\textbf {\bibinfo {volume} {108}},\ \href {https://doi.org/10.1103/physrevd.108.072007} {10.1103/physrevd.108.072007} (\bibinfo {year} {2023})\BibitemShut {NoStop}%
\bibitem [{\citenamefont {Gong}\ \emph {et~al.}(2022)\citenamefont {Gong}, \citenamefont {Meng}, \citenamefont {Zhang}, \citenamefont {Qu}, \citenamefont {Li}, \citenamefont {Qian}, \citenamefont {Du}, \citenamefont {Ma},\ and\ \citenamefont {Liu}}]{jettag2}%
  \BibitemOpen
  \bibfield  {author} {\bibinfo {author} {\bibfnamefont {S.}~\bibnamefont {Gong}}, \bibinfo {author} {\bibfnamefont {Q.}~\bibnamefont {Meng}}, \bibinfo {author} {\bibfnamefont {J.}~\bibnamefont {Zhang}}, \bibinfo {author} {\bibfnamefont {H.}~\bibnamefont {Qu}}, \bibinfo {author} {\bibfnamefont {C.}~\bibnamefont {Li}}, \bibinfo {author} {\bibfnamefont {S.}~\bibnamefont {Qian}}, \bibinfo {author} {\bibfnamefont {W.}~\bibnamefont {Du}}, \bibinfo {author} {\bibfnamefont {Z.-M.}\ \bibnamefont {Ma}},\ and\ \bibinfo {author} {\bibfnamefont {T.-Y.}\ \bibnamefont {Liu}},\ }\bibfield  {title} {\bibinfo {title} {An efficient lorentz equivariant graph neural network for jet tagging},\ }\bibfield  {journal} {\bibinfo  {journal} {Journal of High Energy Physics}\ }\textbf {\bibinfo {volume} {2022}},\ \href {https://doi.org/10.1007/jhep07(2022)030} {10.1007/jhep07(2022)030} (\bibinfo {year} {2022})\BibitemShut {NoStop}%
\bibitem [{GN1(2022)}]{GN1}%
  \BibitemOpen
  \href {https://cds.cern.ch/record/2811135} {\emph {\bibinfo {title} {{Graph Neural Network Jet Flavour Tagging with the ATLAS Detector}}}},\ \bibinfo {type} {Tech. Rep.}\ (\bibinfo  {institution} {CERN},\ \bibinfo {address} {Geneva},\ \bibinfo {year} {2022})\ \bibinfo {note} {all figures including auxiliary figures are available at https://atlas.web.cern.ch/Atlas/GROUPS/PHYSICS/PUBNOTES/ATL-PHYS-PUB-2022-027}\BibitemShut {NoStop}%
\bibitem [{\citenamefont {Pata}\ \emph {et~al.}(2023)\citenamefont {Pata}, \citenamefont {Duarte}, \citenamefont {Mokhtar}, \citenamefont {Wulff}, \citenamefont {Yoo}, \citenamefont {Vlimant}, \citenamefont {Pierini},\ and\ \citenamefont {Girone}}]{CMSpflow}%
  \BibitemOpen
  \bibfield  {author} {\bibinfo {author} {\bibfnamefont {J.}~\bibnamefont {Pata}}, \bibinfo {author} {\bibfnamefont {J.}~\bibnamefont {Duarte}}, \bibinfo {author} {\bibfnamefont {F.}~\bibnamefont {Mokhtar}}, \bibinfo {author} {\bibfnamefont {E.}~\bibnamefont {Wulff}}, \bibinfo {author} {\bibfnamefont {J.}~\bibnamefont {Yoo}}, \bibinfo {author} {\bibfnamefont {J.-R.}\ \bibnamefont {Vlimant}}, \bibinfo {author} {\bibfnamefont {M.}~\bibnamefont {Pierini}},\ and\ \bibinfo {author} {\bibfnamefont {M.}~\bibnamefont {Girone}},\ }\bibfield  {title} {\bibinfo {title} {Machine learning for particle flow reconstruction at {CMS}},\ }\href {https://doi.org/10.1088/1742-6596/2438/1/012100} {\bibfield  {journal} {\bibinfo  {journal} {Journal of Physics: Conference Series}\ }\textbf {\bibinfo {volume} {2438}},\ \bibinfo {pages} {012100} (\bibinfo {year} {2023})}\BibitemShut {NoStop}%
\bibitem [{\citenamefont {Pata}\ \emph {et~al.}(2021)\citenamefont {Pata}, \citenamefont {Duarte}, \citenamefont {Vlimant}, \citenamefont {Pierini},\ and\ \citenamefont {Spiropulu}}]{MLPF}%
  \BibitemOpen
  \bibfield  {author} {\bibinfo {author} {\bibfnamefont {J.}~\bibnamefont {Pata}}, \bibinfo {author} {\bibfnamefont {J.}~\bibnamefont {Duarte}}, \bibinfo {author} {\bibfnamefont {J.-R.}\ \bibnamefont {Vlimant}}, \bibinfo {author} {\bibfnamefont {M.}~\bibnamefont {Pierini}},\ and\ \bibinfo {author} {\bibfnamefont {M.}~\bibnamefont {Spiropulu}},\ }\bibfield  {title} {\bibinfo {title} {{MLPF}: efficient machine-learned particle-flow reconstruction using graph neural networks},\ }\bibfield  {journal} {\bibinfo  {journal} {The European Physical Journal C}\ }\textbf {\bibinfo {volume} {81}},\ \href {https://doi.org/10.1140/epjc/s10052-021-09158-w} {10.1140/epjc/s10052-021-09158-w} (\bibinfo {year} {2021})\BibitemShut {NoStop}%
\bibitem [{\citenamefont {Di~Bello}\ \emph {et~al.}(2023)\citenamefont {Di~Bello}, \citenamefont {Dreyer}, \citenamefont {Ganguly}, \citenamefont {Gross}, \citenamefont {Heinrich}, \citenamefont {Ivina}, \citenamefont {Kado}, \citenamefont {Kakati}, \citenamefont {Santi}, \citenamefont {Shlomi},\ and\ \citenamefont {Tusoni}}]{HGpflow}%
  \BibitemOpen
  \bibfield  {author} {\bibinfo {author} {\bibfnamefont {F.~A.}\ \bibnamefont {Di~Bello}}, \bibinfo {author} {\bibfnamefont {E.}~\bibnamefont {Dreyer}}, \bibinfo {author} {\bibfnamefont {S.}~\bibnamefont {Ganguly}}, \bibinfo {author} {\bibfnamefont {E.}~\bibnamefont {Gross}}, \bibinfo {author} {\bibfnamefont {L.}~\bibnamefont {Heinrich}}, \bibinfo {author} {\bibfnamefont {A.}~\bibnamefont {Ivina}}, \bibinfo {author} {\bibfnamefont {M.}~\bibnamefont {Kado}}, \bibinfo {author} {\bibfnamefont {N.}~\bibnamefont {Kakati}}, \bibinfo {author} {\bibfnamefont {L.}~\bibnamefont {Santi}}, \bibinfo {author} {\bibfnamefont {J.}~\bibnamefont {Shlomi}},\ and\ \bibinfo {author} {\bibfnamefont {M.}~\bibnamefont {Tusoni}},\ }\bibfield  {title} {\bibinfo {title} {Reconstructing particles in jets using set transformer and hypergraph prediction networks},\ }\bibfield  {journal} {\bibinfo  {journal} {The European Physical Journal C}\ }\textbf {\bibinfo {volume} {83}},\ \href {https://doi.org/10.1140/epjc/s10052-023-11677-7}
  {10.1140/epjc/s10052-023-11677-7} (\bibinfo {year} {2023})\BibitemShut {NoStop}%
\bibitem [{\citenamefont {Liu}\ \emph {et~al.}(2023{\natexlab{b}})\citenamefont {Liu}, \citenamefont {Fan}, \citenamefont {Liu}, \citenamefont {Li}, \citenamefont {Li}, \citenamefont {Liu}, \citenamefont {Tang},\ and\ \citenamefont {Li}}]{graphdiffreview}%
  \BibitemOpen
  \bibfield  {author} {\bibinfo {author} {\bibfnamefont {C.}~\bibnamefont {Liu}}, \bibinfo {author} {\bibfnamefont {W.}~\bibnamefont {Fan}}, \bibinfo {author} {\bibfnamefont {Y.}~\bibnamefont {Liu}}, \bibinfo {author} {\bibfnamefont {J.}~\bibnamefont {Li}}, \bibinfo {author} {\bibfnamefont {H.}~\bibnamefont {Li}}, \bibinfo {author} {\bibfnamefont {H.}~\bibnamefont {Liu}}, \bibinfo {author} {\bibfnamefont {J.}~\bibnamefont {Tang}},\ and\ \bibinfo {author} {\bibfnamefont {Q.}~\bibnamefont {Li}},\ }\bibfield  {title} {\bibinfo {title} {Generative diffusion models on graphs: Methods and applications},\ }in\ \href {https://doi.org/10.24963/ijcai.2023/751} {\emph {\bibinfo {booktitle} {Proceedings of the Thirty-Second International Joint Conference on Artificial Intelligence, {IJCAI-23}}}}\ (\bibinfo  {publisher} {International Joint Conferences on Artificial Intelligence Organization},\ \bibinfo {year} {2023})\ pp.\ \bibinfo {pages} {6702--6711}\BibitemShut {NoStop}%
\bibitem [{\citenamefont {Xu}\ \emph {et~al.}(2023)\citenamefont {Xu}, \citenamefont {Powers}, \citenamefont {Dror}, \citenamefont {Ermon},\ and\ \citenamefont {Leskovec}}]{molecules}%
  \BibitemOpen
  \bibfield  {author} {\bibinfo {author} {\bibfnamefont {M.}~\bibnamefont {Xu}}, \bibinfo {author} {\bibfnamefont {A.}~\bibnamefont {Powers}}, \bibinfo {author} {\bibfnamefont {R.}~\bibnamefont {Dror}}, \bibinfo {author} {\bibfnamefont {S.}~\bibnamefont {Ermon}},\ and\ \bibinfo {author} {\bibfnamefont {J.}~\bibnamefont {Leskovec}},\ }\href@noop {} {\bibinfo {title} {Geometric latent diffusion models for 3d molecule generation}} (\bibinfo {year} {2023}),\ \Eprint {https://arxiv.org/abs/2305.01140} {arXiv:2305.01140 [cs.LG]} \BibitemShut {NoStop}%
\bibitem [{\citenamefont {Song}\ \emph {et~al.}(2020)\citenamefont {Song}, \citenamefont {Meng},\ and\ \citenamefont {Ermon}}]{ddim}%
  \BibitemOpen
  \bibfield  {author} {\bibinfo {author} {\bibfnamefont {J.}~\bibnamefont {Song}}, \bibinfo {author} {\bibfnamefont {C.}~\bibnamefont {Meng}},\ and\ \bibinfo {author} {\bibfnamefont {S.}~\bibnamefont {Ermon}},\ }\bibfield  {title} {\bibinfo {title} {Denoising diffusion implicit models},\ }\href@noop {} {\bibfield  {journal} {\bibinfo  {journal} {CoRR}\ }\textbf {\bibinfo {volume} {abs/2010.02502}} (\bibinfo {year} {2020})},\ \Eprint {https://arxiv.org/abs/2010.02502} {2010.02502} \BibitemShut {NoStop}%
\bibitem [{\citenamefont {Nichol}\ and\ \citenamefont {Dhariwal}(2021)}]{improved_diffu}%
  \BibitemOpen
  \bibfield  {author} {\bibinfo {author} {\bibfnamefont {A.~Q.}\ \bibnamefont {Nichol}}\ and\ \bibinfo {author} {\bibfnamefont {P.}~\bibnamefont {Dhariwal}},\ }\bibfield  {title} {\bibinfo {title} {Improved denoising diffusion probabilistic models},\ }in\ \href {https://proceedings.mlr.press/v139/nichol21a.html} {\emph {\bibinfo {booktitle} {Proceedings of the 38th International Conference on Machine Learning}}},\ \bibinfo {series} {Proceedings of Machine Learning Research}, Vol.\ \bibinfo {volume} {139}\ (\bibinfo  {publisher} {PMLR},\ \bibinfo {year} {2021})\ p.\ \bibinfo {pages} {8162},\ \Eprint {https://arxiv.org/abs/2102.09672} {arXiv:2102.09672 [cs.LG]} \BibitemShut {NoStop}%
\bibitem [{\citenamefont {Liu}\ \emph {et~al.}(2022)\citenamefont {Liu}, \citenamefont {Ren}, \citenamefont {Lin},\ and\ \citenamefont {Zhao}}]{liu_pseudo_2022}%
  \BibitemOpen
  \bibfield  {author} {\bibinfo {author} {\bibfnamefont {L.}~\bibnamefont {Liu}}, \bibinfo {author} {\bibfnamefont {Y.}~\bibnamefont {Ren}}, \bibinfo {author} {\bibfnamefont {Z.}~\bibnamefont {Lin}},\ and\ \bibinfo {author} {\bibfnamefont {Z.}~\bibnamefont {Zhao}},\ }\href@noop {} {\bibinfo {title} {Pseudo {Numerical} {Methods} for {Diffusion} {Models} on {Manifolds}}} (\bibinfo {year} {2022}),\ \Eprint {https://arxiv.org/abs/2202.09778} {arXiv:2202.09778} \BibitemShut {NoStop}%
\bibitem [{\citenamefont {Gilmer}\ \emph {et~al.}(2017)\citenamefont {Gilmer}, \citenamefont {Schoenholz}, \citenamefont {Riley}, \citenamefont {Vinyals},\ and\ \citenamefont {Dahl}}]{mpnn}%
  \BibitemOpen
  \bibfield  {author} {\bibinfo {author} {\bibfnamefont {J.}~\bibnamefont {Gilmer}}, \bibinfo {author} {\bibfnamefont {S.~S.}\ \bibnamefont {Schoenholz}}, \bibinfo {author} {\bibfnamefont {P.~F.}\ \bibnamefont {Riley}}, \bibinfo {author} {\bibfnamefont {O.}~\bibnamefont {Vinyals}},\ and\ \bibinfo {author} {\bibfnamefont {G.~E.}\ \bibnamefont {Dahl}},\ }\href@noop {} {\bibinfo {title} {Neural message passing for quantum chemistry}} (\bibinfo {year} {2017}),\ \Eprint {https://arxiv.org/abs/1704.01212} {arXiv:1704.01212 [cs.LG]} \BibitemShut {NoStop}%
\bibitem [{\citenamefont {Wang}\ \emph {et~al.}(2020)\citenamefont {Wang}, \citenamefont {Zheng}, \citenamefont {Ye}, \citenamefont {Gan}, \citenamefont {Li}, \citenamefont {Song}, \citenamefont {Zhou}, \citenamefont {Ma}, \citenamefont {Yu}, \citenamefont {Gai}, \citenamefont {Xiao}, \citenamefont {He}, \citenamefont {Karypis}, \citenamefont {Li},\ and\ \citenamefont {Zhang}}]{dgl}%
  \BibitemOpen
  \bibfield  {author} {\bibinfo {author} {\bibfnamefont {M.}~\bibnamefont {Wang}}, \bibinfo {author} {\bibfnamefont {D.}~\bibnamefont {Zheng}}, \bibinfo {author} {\bibfnamefont {Z.}~\bibnamefont {Ye}}, \bibinfo {author} {\bibfnamefont {Q.}~\bibnamefont {Gan}}, \bibinfo {author} {\bibfnamefont {M.}~\bibnamefont {Li}}, \bibinfo {author} {\bibfnamefont {X.}~\bibnamefont {Song}}, \bibinfo {author} {\bibfnamefont {J.}~\bibnamefont {Zhou}}, \bibinfo {author} {\bibfnamefont {C.}~\bibnamefont {Ma}}, \bibinfo {author} {\bibfnamefont {L.}~\bibnamefont {Yu}}, \bibinfo {author} {\bibfnamefont {Y.}~\bibnamefont {Gai}}, \bibinfo {author} {\bibfnamefont {T.}~\bibnamefont {Xiao}}, \bibinfo {author} {\bibfnamefont {T.}~\bibnamefont {He}}, \bibinfo {author} {\bibfnamefont {G.}~\bibnamefont {Karypis}}, \bibinfo {author} {\bibfnamefont {J.}~\bibnamefont {Li}},\ and\ \bibinfo {author} {\bibfnamefont {Z.}~\bibnamefont {Zhang}},\ }\href@noop {} {\bibinfo {title} {Deep graph library: A graph-centric, highly-performant package for
  graph neural networks}} (\bibinfo {year} {2020}),\ \Eprint {https://arxiv.org/abs/1909.01315} {arXiv:1909.01315 [cs.LG]} \BibitemShut {NoStop}%
\bibitem [{\citenamefont {Jiang}\ \emph {et~al.}(2024)\citenamefont {Jiang}, \citenamefont {Qian},\ and\ \citenamefont {Qu}}]{chooseyourdiffusion}%
  \BibitemOpen
  \bibfield  {author} {\bibinfo {author} {\bibfnamefont {C.}~\bibnamefont {Jiang}}, \bibinfo {author} {\bibfnamefont {S.}~\bibnamefont {Qian}},\ and\ \bibinfo {author} {\bibfnamefont {H.}~\bibnamefont {Qu}},\ }\bibfield  {title} {\bibinfo {title} {{Choose Your Diffusion: Efficient and flexible ways to accelerate the diffusion model in fast high energy physics simulation}},\ }\href@noop {} {\  (\bibinfo {year} {2024})},\ \Eprint {https://arxiv.org/abs/2401.13162} {arXiv:2401.13162 [physics.ins-det]} \BibitemShut {NoStop}%
\end{thebibliography}%


%
\end{document}